\documentclass[twocolumn]{openjournal}

\usepackage{lipsum}

\usepackage{xcolor}
\usepackage{textgreek}
\usepackage[utf8]{inputenc}
\usepackage[english]{babel}
\usepackage{newtxtext,newtxmath}
\usepackage{soul}

\usepackage{hyperref}
\hypersetup{
    unicode, 
    colorlinks=true,
    linkcolor=linkcolor,
    citecolor=linkcolor,
    filecolor=linkcolor,
    urlcolor=linkcolor,
}
\usepackage{color,colortbl}
\definecolor{linkcolor}{rgb}{0.0,0.3,0.5}

\usepackage{tensind}
\tensordelimiter{?}
\DeclareGraphicsExtensions{.bmp,.png,.jpg,.pdf}
\usepackage{verbatim}
\usepackage[normalem]{ulem}
\usepackage{orcidlink}
\definecolor{blazeorange}{rgb}{1.0, 0.4, 0.0}
\definecolor{seagreen}{rgb}{0.18, 0.55, 0.34}
\definecolor{rufous}{rgb}{0.66, 0.11, 0.03}
\definecolor{royalfuchsia}{rgb}{0.79, 0.17, 0.57}
\definecolor{scarlet}{rgb}{1.0, 0.13, 0.0}
\definecolor{royalpurple}{rgb}{0.47, 0.32, 0.66}
\definecolor{darkblue}{rgb}{0, 0, 0.66}

\graphicspath{ {./figs/} }

\begin{document}
\title{Ultra-long Gamma-ray Bursts from Micro-Tidal Disruption Events: The Case of GRB 250702B}

\author{Paz Beniamini\orcidlink{0000-0001-7833-1043}}
\email{pazb@openu.ac.il}
\affiliation{Department of Natural Sciences, The Open University of Israel, P.O Box 808, Ra'anana 4353701, Israel}
\affiliation{Astrophysics Research Center of the Open University (ARCO), The Open University of Israel, P.O Box 808, Ra'anana 4353701, Israel}
\affiliation{Department of Physics, The George Washington University, 725 21st Street NW, Washington, DC 20052, USA}

\author{Hagai B. Perets}
\affiliation{Technion - Israel Institute of Technology, Department of Physics, Technion city, Haifa, Israel 3200002}
\affiliation{Astrophysics Research Center of the Open University (ARCO), The Open University of Israel, P.O Box 808, Ra'anana 4353701, Israel}

\author{Jonathan Granot\orcidlink{0000-0001-8530-8941}}
\affiliation{Department of Natural Sciences, The Open University of Israel, P.O Box 808, Ra'anana 4353701, Israel}
\affiliation{Astrophysics Research Center of the Open University (ARCO), The Open University of Israel, P.O Box 808, Ra'anana 4353701, Israel}
\affiliation{Department of Physics, The George Washington University, 725 21st Street NW, Washington, DC 20052, USA}

\begin{abstract}
    Ultra-long gamma-ray bursts (ULGRBs), a rare class of high-energy transients with durations $>10^3$s, remain poorly understood. GRB 250702B is notable for its multi-hour prompt emission, an X-ray pre-peak emission starting $\sim$1 day earlier, off-nuclear host position, and hard, rapidly variable gamma-rays. This combination is difficult to explain with standard ULGRB progenitors such as blue-supergiant collapsars, magnetar engines, or white-dwarf tidal disruptions by intermediate-mass black holes. We interpret the event as a micro-tidal disruption event ($\mu$TDE), where a stellar-mass black hole or neutron star partially or fully disrupts a main-sequence star. Three $\mu$TDE pathways can reproduce the observed pre-peak emission to main flare delay: (i) a \emph{dynamical (partial/repeating)} disruption, in which a grazing passage yields a faint precursor and the core returns after $\sim$day for a deeper encounter; (ii) a \emph{natal-kick} disruption, where the delay reflects the ballistic motion of a newborn compact object relative to its companion, leading to full disruption; and (iii) a \emph{hybrid natal-kick + partial} case, in which the kick seeds the close encounter but the first passage is only partial, with the core returning on the day-scale period. Cross-section scalings imply comparable rates for partial and full outcomes in both dynamical and natal-kick scenarios. The highly variable, hard $\gamma$-ray emission supports association with a stellar-mass compact object. Fallback and viscous accretion naturally explain the ultra-long duration, energetics, and ks-scale X-ray variability. We outline observational discriminants between the three channels and argue that $\mu$TDEs offer a compelling framework for ULGRBs such as GRB 250702B.
\end{abstract}


\maketitle

\section{Introduction}

Gamma-ray bursts (GRBs) represent the most luminous explosions in the Universe, and are generally divided into short and long classes associated with compact-object mergers and the collapse of massive stars, respectively \citep{Kouveliotou1993,Woosley1993,Eichler1989}. A small subset of events with durations exceeding $\sim 10^3$~s  has emerged in recent years, the so-called \textit{ultra-long GRBs} (ULGRBs). Prototypes include GRB\,101225A, GRB\,111209A, and GRB\,130925A \citep{2005AstL...31..291T,Gendre2013,Greiner2015,Levan2014}, with additional candidates identified more recently \citep[e.g.,][for an overview]{Gendre2025}. Their extreme durations, spectral characteristics and host galaxies suggest that they are a distinct class from long GRBs \citep{Levan2014,Boer2015} and call for a different astrophysical channel. Mixed evidence regarding accompanying supernovae, spectral diversity, and lightcurve morphology within the ULGRB class could potentially indicate that they are not a homogeneous group.

Another class of ultra-long duration X-ray/gamma-ray events, including Sw1644+57, J2058+05, J1112-8238 and AT2022CMC, were suggested to originate from stellar disruptions by intermediate (IMBH) or supermassive (SMBH) black holes. They are observationally distinct from ULGRBs in several ways: (i) whenever evidence is available, they appear to coincide with their galactic centers\footnote{The first likely identification of an off-center TDE, TDE2024tvd was detected in the optical band by ZTF \citep{Yao2025}. It exhibits peculiar radio properties which appears to be inconsistent with having produced a relativistic outflow \citep{Sfaradi2025}.}, (ii) they have longer durations (days to weeks), (iii) they have fainter peak luminosities ($L_{\rm X,iso}\sim 10^{47}-10^{48}\mbox{erg s}^{-1}$), (iv) their emission is soft, typically peaking at a few tens of keV, (v) they have long variability times ($\sim 10^2-10^3$\,s), (vi) in the cases where X-rays could be monitored for months after the initial trigger, an abrupt shut-off has been observed on that timescale (commonly interpreted as a change in the accretion mode of the disk) and (vii) they have no associated SNe.

Several models have been proposed to explain ULGRBs. The collapse of blue supergiants can, in principle, provide long-lived accretion due to the extended stellar envelope \citep{Woosley2012,Levan2014}, but should generally be accompanied by bright hydrogen-rich supernovae (although see \citealt{Perna2018} for a blue supergiant from a helium rich star). Newly born magnetar `engines' can inject sustained power into relativistic outflows \citep{Metzger2015,Gompertz2017}, but rarely predict long duration pre-peak emission starting many hours before the main flare. Mergers of WDs with other compact objects have also been proposed \citep{Fryer2019}, although the rate estimates are typically low, and it is unclear whether jets can robustly form in these environments \citep{Narayan2001}. Micro tidal disruptions ($\mu$TDEs; \cite{Perets2016}), i.e., the disruption of stars/planets by stellar compact objects, have been suggested as a channel for the production of ultra-long-duration GRBs. None of the models has yet gained sufficient support to explain the full phenomenology of ULGRBs, and more than a single channel might be involved in producing different ULGRBs.

We consider a ULGRB candidate, GRB 250702B, displaying remarkable features including an exceptionally long duration of several hours with variability on timescales as short as $\sim0.5$\,s and $\gtrsim 10$MeV photons (in the source's cosmological frame), and a soft pre-peak signal initially detected in X-rays nearly one day prior to the main burst. The event occurred significantly offset from its host-galaxy nucleus, and no accompanying supernova was detected, although the redshift and dust extinction likely preclude very stringent limits. These properties are difficult to reconcile with the majority of ULGRB models. 

We explore the possibility that this event was powered by a $\mu$TDE, in this case the disruption or partial stripping of a main-sequence star by a stellar-mass black hole or neutron star. 
This channel was first introduced by \citet{Perets2016}, who identified three formation routes: random encounters in dense clusters, tidal capture or scattering in the field, and close passages triggered by natal kicks of newborn compact objects. Hydrodynamic simulations and analytic work have since established that such events can produce long-duration high-energy transients with fallback rates consistent with ULGRBs \citep[e.g.,][]{Wang2021,Xin2024,Vynatheya2024}. Repeated or partial disruptions can give rise to multiple flares separated by the orbital period of the surviving stellar core \citep{Coughlin2019,Wang2021}, while a full disruption by natal-kick can naturally generate delays of hours to days between an initial explosive transient and/or between the SN ejecta - companion collision and the subsequent disruption \citep{Michaely2016,Perets2016,BP2024}. 

In this work, we analyze the observed properties of the event in the context of the $\mu$TDE hypothesis. We show that $\mu$TDEs can be considered as `missing links' between relativistic TDEs and collapsars or mergers involving stellar mass compact objects. The long duration and the day-earlier emission start-time in X-rays can be readily explained in this framework, while still allowing for `GRB-like' features, such as rapid variability, high photon energies, large prompt luminosities, and an off-center galactic location. We compare to alternative models such as collapsars, magnetars, and white-dwarf IMBH TDEs. We argue that this event may represent the first substantial evidence for a $\mu$TDE origin of an ultra-long GRB.

\section{Observational Properties}
The burst was discovered on 2025 July 2 by several facilities, including {\it Fermi}-GBM, Konus-{\it Wind}, {\it SVOM}, {\it MAXI}, and {\it Einstein Probe}, with a total source-frame prompt duration of $\approx 4.2$ hours in the $\gamma$-ray bands \citep{Levan2025,Oganesyan2025,Neights2025,O'Connor2025,Gompertz2025,Carney2025,SVOM,MAXI25GCN.40910,konusgcn}. The time-integrated fluence detected by Fermi-GBM and Konus-Wind was $\sim 4.6\times 10^{-4}\ \mathrm{erg\ cm^{-2}}$, and the prompt spectrum is well described by a hard cut-off power-law $\gamma$-ray spectrum. Astrometry places the transient significantly offset from the host galaxy nucleus by $\sim 5.7$\,kpc (projected; \citealt{Levan2025,Carney2025}).
The inferred stellar mass of the host galaxy is large, $\sim 10^{10.5}M_{\odot}$, and highly asymmetric - indicative of a potential recent galaxy merger.
No accompanying supernova was identified in late-time optical observations; at $z\simeq 1.04$ \citep{Gompertz2025} and considering the significant amount of dust extinction, this non-detection primarily rules out only the most luminous events and does not impose stringent constraints on ordinary SNe.

A notable feature is a soft X-ray ($0.5-4\;$keV) signal initially detected approximately one day prior to the main burst \citep{EP25GCN.40906,EP250702a}. This signal was fainter than the main episode, and exhibited a softer spectrum than the subsequent prompt emission. At the time of writing it is not yet clear if there was an emission gap between the earliest X-ray detections and the main $\gamma$-ray event. While late-time X-ray flares and extended activity are known in some GRBs, a $\sim$day-earlier signal preceding an ultra-long, high-energy episode has not been clearly documented before; importantly, previous ULGRBs predated {\it Einstein Probe}, so similar pre-peak signals could have been missed. Nevertheless, one should note the case of GRB 050709 where a short-GRB was identified 16 days before a two-hour-long duration X-ray flare \citep{Fox2005} (although the duration and fluence ratio between the preceding emission and the following event were very different than those observed in GRB 250702).

The key measured properties (quoted in the source frame unless noted) are as follows. The prompt emission is ultra-long, lasting $T_{90}\gtrsim 12$\,ks, yet shows sub-second structure with a minimum variability time of $\delta t_{\rm rest}\!\sim\!0.5$\,s and photons extending to $\gtrsim 10$\,MeV, implying a large bulk Lorentz factor from standard compactness arguments \citep{Neights2025}.
We note that while the minimum variability timescale (MVT) is limited by photon statistics and thus has moderate uncertainty, $\delta t_{\rm obs}=1\pm 0.4$, a similar MVT is found by \cite{Neights2025} in three independent time intervals and is supported by analysis of Psyche-GRNS data in addition to Fermi-GBM. We therefore interpret the minimum variability time as an order-of-magnitude constraint, indicating characteristic variability on a rest frame timescale of $\sim 0.5-\text{few}$\,s, rather than $\gtrsim 10$\,s. Crucially, this is sufficient to position GRB 250702B apart from relativistic TDEs such as Swift J1644+57
An X-ray pre-peak emission component occurred $\sim 0.5$\,day before the main episode \citep{EP250702a}. The transient is offset from the irregular galaxy's host nucleus and no luminous supernova was detected. 
Late-time emission from the event has been observed across multiple wavelengths, from radio \citep{2025GCN.41059....1A,2025GCN.41147....1G,2025GCN.41145....1B,2025GCN.41061....1T,2025GCN.41053....1S,2025GCN.41054....1A} to IR \citep{Carney2025,Levan2025} and X-rays \citep{O'Connor2025}. The broadband nature of the emission favors a non-thermal origin.
The $0.3$--$10$\,keV light curve decays as $\sim t^{-1.8}$ when anchored at the brightest GBM trigger (E) and exhibits superposed variability on $\sim 0.5$--$1$\,ks scales within the first few days \citep{O'Connor2025}. 
Broadband afterglow modeling favors a wind-like external medium ($n\!\propto\! r^{-2}$) and possibly a narrow, ultra-relativistic jet \citep{O'Connor2025,Levan2025}, although large degeneracy remains considering the observations to date. The ratio between the isotropic prompt $\gamma$-ray energy and the X-ray luminosity at 11\,hr is $E_{\gamma,54}/L_{X,11\,\mathrm{hr},47}\!\sim\!1$ \citep{O'Connor2025}, consistent with the broader GRB population, including that of ULGRBs. We note that the choice of zero time $t_0$ affects the estimate of $L_{\rm X,11 hr}$ as well as the early temporal indices: taking $t_0$ at the initial X-ray pre-peak emission naturally steepens the apparent initial decay before relaxing toward the canonical fallback-like behavior (see §\ref{subsec:partial}; see also \citealt{Levan2025}, for observational details).

\section{The Micro-TDE Scenario}
We consider three distinct pathways capable of producing a $\sim$day-earlier pre-peak component and an ultra-long main flare: (i) a \emph{dynamical (partial/repeating)} $\mu$TDE, where a first grazing passage strips a small amount of mass and the surviving core returns on an orbital timescale for a deeper encounter; (ii) a \emph{natal-kick} $\mu$TDE, where a newborn compact object receives a kick and, after a ballistic transit, passes sufficiently close to its companion to cause a disruption; and (iii) a \emph{hybrid natal-kick + partial} case, where the kick both creates the close orbit and the first peri-center passage is partial, so the surviving core returns after $\sim$day for the dominant flare. Sections~\S\ref{sec:TDEbasic}--\S\ref{sec:Fallback} lay out basic disruption, timescale, and energetic scalings; \S\ref{subsec:partial} treats the partial/repeating channel (potentially due to dynamical interactions), \S\ref{subsec:Kicks} the natal-kick channel, and \S\ref{subsec:hybrid} the hybrid case. Section \S\ref{subsec:afterglow} discusses implications for afterglow emission properties.

\subsection{Tidal Disruption Basics and energetics}
\label{sec:TDEbasic}
The defining scale of a tidal disruption event is the tidal radius,
at which the star is torn apart by the differential gravitational 
forces of the compact object. For a star of mass $M_\ast$ and radius $R_\ast$, and a compact object of mass $M_\bullet$, this is
\begin{equation}
\label{eq:rT}
    r_{\rm t} \;\approx\; 1.5\times 10^{11}\,\eta^{2/3}\,
    M_{\bullet,1}^{1/3}\,M_{\ast}^{-1/3}\,R_{\ast}
    \ {\rm cm}\;,
\end{equation}
where $M_{\bullet,1}=M_\bullet/10\,M_\odot$, $M_\ast$ and $R_\ast$ are in solar units,
and $\eta \sim 1$ is a dimensionless parameter encapsulating the internal density profile of the star and is of order unity for main-sequence stars (see, e.g., \citealt{Phinney1989,Lodato2009}). We focus here on $M_{\bullet}>M_{\ast}$, such that $r_t\gtrsim R_{\ast}$. This is in order to avoid a `direct collision', which would likely result in a very rapid asymmetric disruption with much of the material swallowed whole by the black hole rather than forming a bound debris stream as expected in classic TDEs. For this reason, we also assume below that $M_{\bullet}$ dominates the final bound mass of the system, which at worst corresponds to an order unity approximation.
For a solar-type star disrupted by a stellar-mass black hole, $r_{\rm t}$ is of order a few solar radii, comfortably larger than the Schwarzschild radius ($\sim 30$ km for $10\,M_\odot$), ensuring that full disruption is possible.

A close passage with peri-center distance $r_p \lesssim r_{\rm t}$ leads to significant mass loss, with the disruption strength commonly parameterized by the penetration factor $\beta \equiv r_{\rm t}/r_p$. Encounters with $\beta \gtrsim 1$ typically yield complete disruptions, while shallower encounters ($\beta \lesssim 1$) can produce partial disruptions and repeated stripping events.
This distinction is particularly important for interpreting the pre-peak X-ray emission in GRB 250702B: a grazing first passage can strip a small fraction of the stellar envelope, producing an initial faint flare. This could lead to effective tidal capture of a companion, even if the disrupted star was not originally bound to the system. In any case, the subsequent peri-center return of the surviving core on an orbital timescale can generate the dominant ultra-long burst.

\subsection{Fallback and Accretion: Timescales and Energetics}
\label{sec:Fallback}
Once a star is disrupted, approximately half of the stellar material becomes
gravitationally bound and returns to the compact object on a range of fallback
timescales. The earliest returning debris sets the onset of accretion at a time
comparable to the orbital time of the most bound material,
\begin{equation}
\label{eq:tmin}
    t_{\rm min} \;\approx\; 1.1\times 10^4 \,
    \eta^2 \, M_{\bullet,1}^{1/2}\, R_\ast^{3/2} \, M_\ast^{-1}A_{\beta}
    \ {\rm s}\;,
\end{equation}
where $A_{\beta}$ is a parameter that depends on the response of the star to the tidal field. In the `frozen in' approximation, disruption occurs at the tidal radius, and hence $A_{\beta}=1$. However, if energy can be dissipated efficiently near the peri-center, then it is the latter that dominates the energy of the most bound material and one obtains $A_{\beta}\approx \beta^{-3}$.  
For a $10\,M_\odot$ black hole disrupting a solar-type star,
$t_{\rm min}$ is of order a 1-2 hours, in line with the expected rise times of ultra-long bursts \citep{Perets2016}. 
The mass fallback rate then declines as $\dot{M}\propto t^{-5/3}$ in the case of complete disruption, 
or steeper for partial disruptions \citep{Coughlin2019,Wang2021}.

The effective accretion duration also depends on the circularization and viscous timescales of the debris disk. 
If the material circularizes near twice the peri-center distance (this is consistent with most of the dissipation occurring near the peri-center), then $R_{\rm circ} \simeq 2 r_p = 2r_{\rm t}/\beta$.
The corresponding viscous inflow timescale is
\begin{equation}
    t_{\rm acc} \;\approx\; 4.5\times 10^4 \,\eta\,\alpha_{-1}^{-1}\,h^{-2}\,\beta^{-3/2}
    M_\ast^{-1/2} R_\ast^{3/2} \ {\rm s}\;,
\end{equation}
where $\alpha=10^{-1}\alpha_{-1}$ is the disk viscosity parameter, $h=H/R$ the disk aspect ratio. Note that $t_{\rm acc}$ is to first order independent of $M_{\bullet}$. For plausible parameters ($\alpha \sim 0.1$, $h \sim 1$, $\beta \sim 1$), $t_{\rm acc}$ is of order $\sim 10^4$--$10^5\;$s (see also Fig.~\ref{fig:Masses_param_space}),  allowing sustained accretion over many hours, as observed.

The condition $t_{\rm min} \lesssim t_{\rm acc}$ ensures that fallback proceeds without stalling, such that the majority of the debris can be efficiently accreted. This is naturally satisfied in the relevant parameter space of stellar-mass black holes and main-sequence stars, further supporting the feasibility of the $\mu$TDE interpretation. There is a critical black hole mass, $M_{\bullet}\approx 200 M_{\ast}\eta^{-2}\alpha_{-1}^{-2}h^{-4}\beta^{-3} A_{\beta}^{-2}M_{\odot}$ above which $t_{\rm min}$ dominates over $t_{\rm acc}$ and the accretion time becomes subdominant. 
This holds for TDEs involving SMBHs, while the two timescales may be comparable for TDEs involving IMBHs (see also \citealt{Granot2025}).

The accretion rate onto the compact object is not identical to the raw fallback rate, but rather is ``viscously filtered'' by the disk response. Writing
\begin{equation}
\dot M_{\rm acc}(t)=\int_{t_{\min}}^t \dot M_{\rm fb}(t')\,\frac{e^{-(t-t')/t_{\rm acc}}}{t_{\rm acc}}\,\mathrm{d}t'\;,
\end{equation}
one finds for a power-law fallback $\dot M_{\rm fb}\propto t^{-n}$ with $n>1$ two regimes:
\begin{align}
t\!\lesssim\!t_{\rm acc}:&\quad \dot M_{\rm acc}(t)\simeq \frac{1}{t_{\rm acc}}\int_{t_{\min}}^t \dot M_{\rm fb}(t')\,\mathrm{d}t' \approx \mathrm{const}\;,\\
t\!\gg\!t_{\rm acc}:&\quad \dot M_{\rm acc}(t)\simeq \dot M_{\rm fb}(t)\propto t^{-n}\;.
\end{align}
Thus, when $t_{\rm acc}\!\gtrsim\!t_{\min}$ a shallow phase of duration $\sim t_{\rm acc}$ precedes the canonical decay. This damps any very-early $t^{-3}$ segment from the weakly bound tail.

In general, the initial shallow phase with the largest accretion rate lasts for $\sim \max(t_{\rm min},t_{\rm acc})$.
These scalings apply equally in both the dynamical and natal-kick channels, setting the duration and energetics of the main flare once debris is supplied to the compact object.

\textbf{Energy budget.} Another consideration regards the overall energetics. The isotropic equivalent $\gamma$-ray energy of GRB 250702B is $\gtrsim 1.4\times 10^{54}\mbox{ erg}$ \citep{Neights2025}. This corresponds to a collimated-corrected kinetic energy of $E_{\rm j}\sim 6.3\times 10^{51}\eta_{\gamma,-1}^{-1}\theta_{\rm j,-1.5}^2\mbox{ erg}$ where $\eta_{\gamma}$ here is the ratio between the $\gamma$-rays and remaining kinetic energy (which is comparable to the $\gamma$-ray efficiency when it is $\ll 1$) and $\theta_{\rm j}$ is the jet opening angle. \cite{Wu2025} have shown that magnetic (Blandford-Znajek) jets operating in short and long GRB engines have a maximal efficiency of $\eta_{\rm j} \sim 0.015$ in converting accretion to jet energy (although higher values are in principle possible for other engines if the black hole mass is large compared to the total accreted mass, as indeed is expected for the canonical parameters for $\mu$TDEs explored here) and found typical values (which in general depend on the magnetic flux) of order $\eta_{\rm j} \sim 10^{-3}$ when comparing to observed GRBs. Similar values of $\eta_{\rm j}$ are obtained using inferred jet energy in relativistic TDE outflows such as Sw1644+57 \citep{BPM2023}.
For GRB 250702B, this estimate of accreted mass implies an accreted mass of
\begin{equation}
\label{eq:Macc}
  M_{\rm acc}\approx 0.35\ \eta_{\gamma,-1}^{-1}\,\theta_{\rm j,-1.5}^2\,\eta_{\rm j,-2}^{-1}\ M_{\odot}\;.
\end{equation}
\cite{Narayan2001} pointed out that, depending on the accretion regime, it is possible that only a fraction, $\eta_{\rm acc}\ll 1$ of mass in the accretion disk ends up falling into the black hole. At the same time, $\eta_{\rm acc}$ cannot be too low, considering the energetics of known GRBs and TDEs and the plausible amounts of mass within their associated accretion disks. Even if the latter is as small as $\eta_{\rm acc}\approx 0.1$, the estimate in Eq. \ref{eq:Macc} is consistent with approximately half of the mass of the disrupted stellar companion, as expected from the $\mu$TDE scenario.

{\bf High energy photons and $\gamma$-ray variability.}
Short observed variability timescales are often interpreted as evidence for a compact central engine; however, this association is not unique. In relativistic jets, variability can in principle be shortened by relativistic motion of the emitting material, either through light-travel-time effects or through relativistic sub-structures moving within the bulk flow. In the absence of such sub-structure, the radius of the causally connected dissipation region increases with the bulk Lorentz factor in a manner that largely compensates for light-travel-time compression, implying that the observed variability remains comparable to the central-engine timescale (e.g. \citealt{Kobayashi1997,BG2016}).

Variability significantly shorter than the central-engine light-crossing time can arise in mini-jet scenarios, which have been invoked to explain rapid TeV variability in blazars (e.g. \citealt{Giannios2009}). However, these models wherein the bulk is patchy and contains emitting regions which move relativistically within it, are characterized by low radiative efficiency \citep{BK2020} and are therefore most naturally applicable to emission components that carry only a small fraction of the total jet energy. Applying such mechanisms to the dominant prompt gamma-ray emission observed in GRB 250702B would require an energetically prohibitive efficiency, leading to an energy-budget problem (as detailed above). We therefore conclude that while relativistic sub-structure may contribute to variability, it is unlikely to substantially relax the variability-based constraints inferred from the prompt emission.

Finally, the detection of $\gtrsim\!10$\,MeV photons with sub-second variability implies, via standard $\gamma\gamma$ opacity arguments with an emission radius $R\!\sim\!\Gamma^2 c\,\delta t$, a bulk Lorentz factor $\Gamma\!\gtrsim\!{\rm few}\times10$--$10^2$ for the prompt emitter \citep[e.g.][]{Neights2025}. This is consistent with a relativistic jet launched by $\mu$TDE accretion and provides an additional consistency check once the final $E_{\rm peak}$ and variability are fixed.

\subsection{Dynamical Channel (partial/repeating): Orbital-Period Delay and Multi-Episode Emission}
\label{subsec:partial}

In a grazing encounter with peri-center $r_p \gtrsim r_{\rm t}$ (penetration factor $\beta \equiv r_{\rm t}/r_p \lesssim 1$), the star suffers \emph{partial} mass loss while a bound stellar core survives.

The stripped debris returns on the fallback timescale (setting an initial, faint pre-peak signal), whereas the surviving core, left on an eccentric orbit about the compact object, returns to peri-center after an orbital period, potentially undergoing a deeper encounter that powers the ultra-long main flare.

Hydrodynamic simulations show that repeated partial disruptions often lead to progressively deeper passages \citep[e.g.,][]{Guillochon2013,Coughlin2019,Wang2021,Kiroglu2023}. The underlying mechanism can be understood analytically (below): the first encounter modifies both the orbital parameters and the stellar structure in ways that reduce the peri-center distance and increase the effective tidal radius.

The orbital period for total post-encounter mass $M_{\rm f}\approx M_{\bullet}$ is
\begin{equation}
    P_{\rm orb} \;=\; 2\pi \left(\frac{a^3}{G M_{\rm f}}\right)^{1/2}
    \;\; \Rightarrow \;\;
    a \;\approx\; 6.3\times 10^{11}\,
    P_{\rm day}^{2/3} M_{\rm \bullet,1}^{1/3}\ {\rm cm}\,,
\end{equation}
where $P_{\rm day}=P_{\rm orb}/{\rm day}$.
To trigger enhanced stripping at the next passage, we require $r_p \sim r_{\rm t}$;
for an eccentric orbit $r_p=a(1-e)$, hence
\begin{equation}
    1-e=\frac{r_{\rm t}}{a}=0.24\,
    \eta^{2/3}\,\frac{ M_\ast^{-1/3} R_\ast}
    {P_{\rm day}^{2/3} }\,, 
\end{equation}
using Eq. \ref{eq:rT}.
For $M_\bullet\!\sim\!10\,M_\odot$, $M_\ast\!\sim\!1\,M_\odot$, $R_\ast\!\sim\!1\,R_\odot$, and $P_{\rm orb}\!\sim\!1$ day,
one finds $a\!\sim\!{\rm few}\times10^{11}$–$10^{12}$ cm and $e\!\sim\!0.7$–$0.9$, placing the next passage again near $r_{\rm t}$.

\textbf{Why is the second passage deeper?}
\emph{(i) Orbital shrinkage from angular-momentum loss.}
Let $(a_0,e_0,\mathscr{l}_0)$ be the pre-passage Keplerian elements and $\mathscr{l}$ the specific angular momentum. For a highly eccentric orbit ($e\!\to\!1$),
\begin{equation}
r_p \simeq \frac{\mathscr{l}^2}{2GM_{\bullet}}\;.
\end{equation}
If the first encounter removes a (small) fraction of angular momentum, $\mathscr{l}_1=\mathscr{l}_0(1-\delta_{\mathscr{l}})$ with $\delta_{\mathscr{l}}>0$, then
\begin{equation}
\frac{r_{p,1}}{r_{p,0}}=\left(\frac{\mathscr{l}_1}{\mathscr{l}_0}\right)^2=(1-\delta_{\mathscr{l}})^2 \simeq 1-2\delta_{\mathscr{l}}\;,
\end{equation}
so $r_{p,1}<r_{p,0}$. Energy dissipation further reduces $a$ but does not enter $r_p$ at leading order in the $e\!\to\!1$ limit. It will, however, reduce the return time as $P_{\rm orb}\propto a^{3/2}$.

\emph{(ii) Stellar inflation and mass loss from partial stripping.}
A grazing pass removes mass and injects entropy, inflating the survivor to $R'_\ast$ and reducing its mass to $M'_\ast<M_\ast$. Since $r_{\rm t}\propto \eta^{2/3}\,M_\bullet^{1/3} M_\ast^{-1/3} R_\ast$, the post-passage tidal radius is
\begin{equation}
\label{eq:rTprime}
\begin{aligned}
r'_{\rm t} \;=\; r_{\rm t}\,
\left(\frac{\eta'}{\eta}\right)^{2/3}
\left(\frac{R'_\ast}{R_\ast}\right)
\left(\frac{M'_\ast}{M_\ast}\right)^{-1/3}\,.
\end{aligned}
\end{equation}
Combining this with the peri-center change from angular-momentum loss,
$r_{p,1}/r_{p,0}=(1-\delta_{\mathscr{l}})^2$, we obtain
\begin{equation}
\label{eq:beta1}
\begin{aligned}
\beta_1 \;\equiv\; \frac{r'_{\rm t}}{r_{p,1}}
&= \beta_0\,
\left(\frac{\eta'}{\eta}\right)^{2/3}
\left(\frac{R'_\ast}{R_\ast}\right)
\left(\frac{M'_\ast}{M_\ast}\right)^{-1/3}
\frac{1}{(1-\delta_{\mathscr{l}})^2}
\\[2pt]
&\simeq\;
\beta_0\!\left[
1
+ \frac{\Delta R_\ast}{R_\ast}
- \frac{1}{3}\frac{\Delta M_\ast}{M_\ast}
+ \frac{2}{3}\frac{\Delta\eta}{\eta}
+ 2\delta_{\mathscr{l}}
\right],
\end{aligned}
\end{equation}
where $\Delta X \equiv X'-X$ and the last line is a first-order expansion. Both the radius increase and the mass decrease raise $\beta$; for $\delta_{\mathscr{l}}>0$ the $2\delta_{\mathscr{l}}$ term is also positive, i.e., the second passage typically has a larger penetration factor, boosting the stripped mass and the fallback peak. 

The energetics and duration of the main episode are then governed by the fallback and viscous timescales derived in \S \ref{sec:Fallback}.
A key diagnostic is the fallback rate:
complete disruptions yield $\dot M \propto t^{-5/3}$ after peak, whereas partial disruptions exhibit a steeper late-time decline,
approaching $\dot M \propto t^{-9/4}$ once a bound core remains \citep[e.g.,][]{Coughlin2019,Wang2021}. Moreover, the accretion behavior in the partial disruption cases could be quite complex and not follow simple power-laws due to clumping and stream collisions \citep{Xin2024}.

Very steep \emph{apparent} declines (up to $\sim t^{-3}$) can occur if the light curve is anchored at the pre-peak signal rather than the main disruption epoch. This “time-zero” effect is distinct from the intrinsic fallback law \citep[e.g.,][]{Guillochon2013,Coughlin2019}. This is naturally expected as (i) the effective zero-time shifts to the second, deeper encounter, and (ii) the disk transitions from circularization-dominated to fallback-dominated accretion.

Spectrally, the pre-peak signal is expected to be softer and less luminous (lower $\beta$, smaller stripped mass), while the second passage, with effectively larger $\beta$ and higher fallback rate, produces the dominant high-energy output. There will be an emission gap between pre-peak signal and the main emission if $\max(t_{\rm acc,0},t_{\rm min,0})<P_{\rm orb}$ (where 0 indicates quantities estimated for the initial grazing passage), while in the opposite limit the observed emission will transition directly from the soft-dim pre-peak signal to the hard-bright main emission.
Finally, because a bound core can survive the second passage if $\beta$ remains below the complete-disruption threshold,
additional, weaker flares are possible on subsequent orbits, with gradually evolving periods as the orbit circularizes, and/or until the final disruptions of the remnant core.
Whether such repeats are detectable depends on how rapidly the core is eroded and on jet collimation/efficiency.

\subsection{Natal-kick Channel: Ballistic Delay and Encounter Probability}
\label{subsec:Kicks}
In the natal-kick scenario, a compact object formed in a core-collapse event receives a kick $\mathbf{v}_{\rm kick}$ relative to its pre-SN motion, changes its orbit/trajectory and encounters the companion to disrupt/partially disrupt it \citep{Perets2016,Michaely2016}. 
The relative speed between the newborn compact object and its companion is then $v' \simeq |\mathbf{v}_{\rm orb}+\mathbf{v}_{\rm kick}|$, where
$v_{\rm orb}\sim (G M_{\rm i}/a_{\rm i})^{1/2}$ is the pre-SN orbital speed and $M_{\rm i}$ and $a_{\rm i}$ are the initial total mass and semi-major axis.

Due to the natal kick, the post-SN orbital parameters become
\[
\epsilon=\frac{v'^2}{2}-\frac{GM_f}{a_i},\qquad
\mathscr{l}=a_i|\hat{\mathbf r}\times \mathbf v'|,
\]
with $\mathbf v'=\mathbf v_{\rm orb}+\mathbf v_{\rm kick}$. 
This directly yields the new semimajor axis $a=-GM_f/(2\epsilon)$ and eccentricity $e=\sqrt{1+2\epsilon \mathscr{l}^2/(G^2M_f^2)}$, so that the peri-center is $r_p=a(1-e)$. 
The time from explosion to the first close passage is then $t_{\rm w}\simeq P_{\rm orb}/2=\pi\sqrt{a^3/(GM_f)}$, which naturally falls in the hour–day regime for $a\sim 10^{11}$–$10^{12}$ cm.

In order for the kick to significantly shorten the peri-center, one must have $a_{\rm i}/2\lesssim a\lesssim a_{\rm i}$ as well as $v_{\rm kick}\approx -v_{\rm orb}$ \citep{Michaely2016,BP2024}, or
\begin{equation}
    v_{\rm kick}\approx 450  \chi^{1/2} M_{\rm f,1}^{1/3} P_{\rm day}^{-1/3} \mbox{km s}^{-1}
\end{equation}
where $\chi\equiv M_{\rm i}/M_{\rm f}\ge1$ is the inverse of the fractional total mass loss. 
The required kick naturally fits within typical kick velocities inferred for NS/BH formation.

\begin{figure*}
    \centering
    \includegraphics[width=0.4\linewidth]{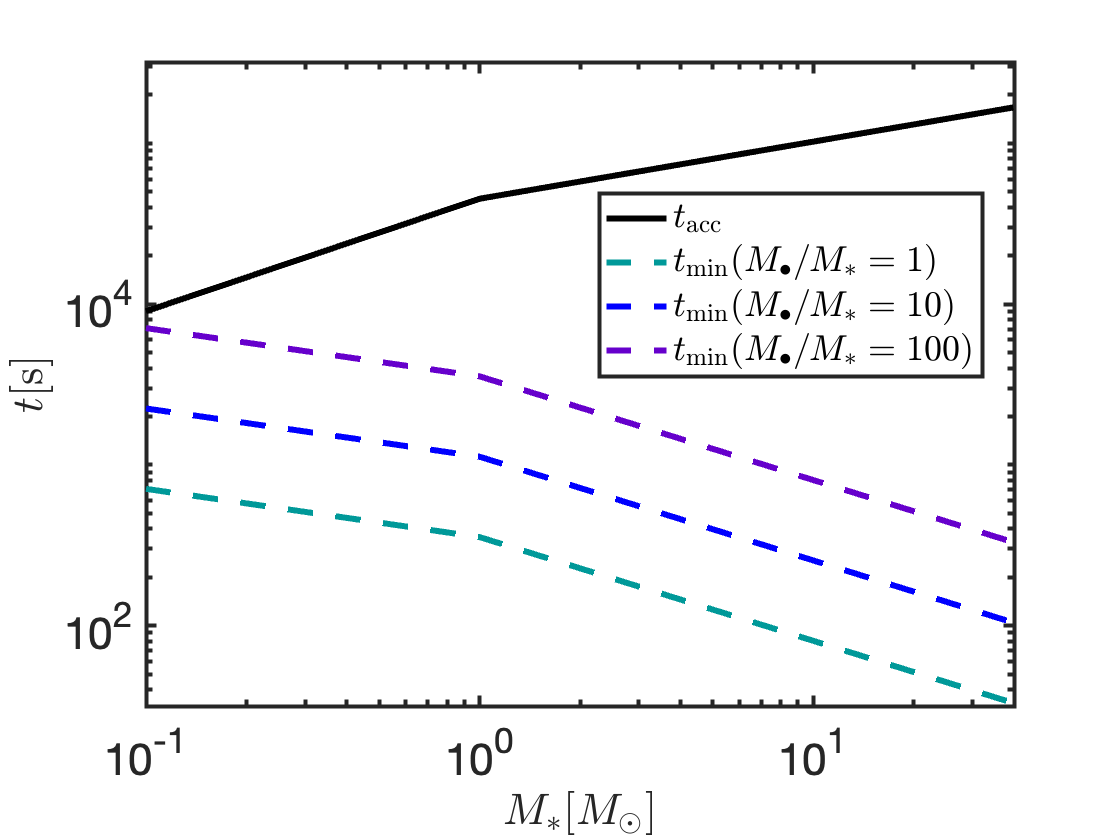}
 \includegraphics[width=0.4\linewidth]{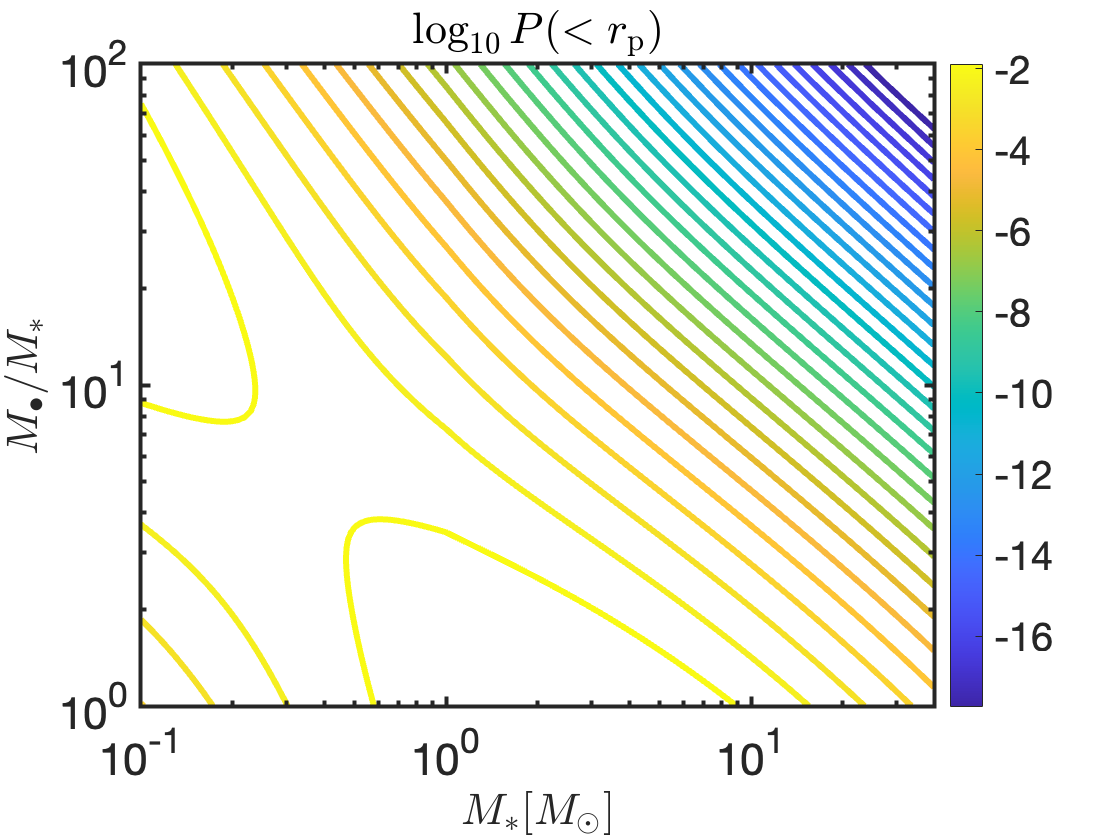}
    \caption{Characteristics of $\mu$TDEs in the for different $M_{\ast}$ and $M_{\bullet}/M_{\ast}$. In all cases we assume main sequence stars (with $R_{\ast}\propto M_{\ast}^{0.8}$ at $M_{\ast}<M_{\odot}$ and $R_{\ast}\propto M_{\ast}^{0.57}$ otherwise). We also assume $\alpha=0.1,h=1,\beta=1,\eta=1,M_{\rm ej}=10^{0.5}M_{\odot},P_{\rm f}=1$\,day (see text for details). Left: Viscous accretion timescale ($t_{\rm acc}$) as compared with the Keplerian timescale of the most bound material ($t_{\rm min}$). Typical values span $10^4<t_{\rm acc}<10^5$\,s. The condition $t_{\rm acc}>t_{\rm min}$ is satisfied in the entire region plotted. Right: Probability that the SN kick results in the final orbit having a peri-center $r_{\rm p}<r_{\rm t}$. We assume `momentum kicks' with a scaling relative to well-studied NS formation kicks, $v_{\rm kick}\approx v_{\rm kick,NS}M_{\rm NS}/M_{\bullet}$ where the NS kick distribution is taken as log-normal with median of $450\mbox{km s}^{-1}$ and $\sigma=0.5$ \citep{Hobbs2005}.}
    \label{fig:Masses_param_space}
\end{figure*}

The probability that the kick sends the system to an orbit with $r_{\rm p}\ll a_i$ is given by \citep{BP2024}
\begin{equation}
    P(<r_p) \;\approx\; \frac{6}{\chi^{3/2}}\,
    P_{v}
    \frac{r_p}{a_{\rm i}} .
\end{equation} where $P_v\equiv P\!\left(v_{\rm orb}\!\lesssim\! v_{\rm kick}\!\lesssim\!2v_{\rm orb}\right)$ is the probability of having the kick aligned within a factor of two of the Keplerian velocity. Setting $r_{\rm p}\approx r_{\rm t}$, this gives a probability of $P(<r_{\rm p})\sim 0.23 P_v$, which is not too small. This is demonstrated in figure \ref{fig:Masses_param_space} in which we plot this probability for different $M_{\bullet},M_{\ast}$. In the majority of the relevant parameter space, $10^{-4}<P(<r_{\rm p})<10^{-1.5}$.

Combining the timing and probability, the natal-kick channel naturally produces (i) a high-energy transient at core collapse, followed $\sim$hours–day later by (ii) a close passage with $r_p\!\lesssim\!r_{\rm t}$ that powers the main, ultra-long flare via fallback accretion. Besides GRB 250702B explored here, such a scenario may also explain the case of GRB 050709, where a kick following a NS merger could have led to a disruption of a companion 16 days later \citep{Perets2016} \footnote{A related scenario is that of a NS merger in a hierarchal triple, that results in a kick that sends the NS merger product to a rapidly merging orbit with a tertiary black hole. This possibility was suggested to explain the large mass ratio black hole merger, GW 190814 detected by LIGO-VIRGO \citep{LBB2021}. The locations/rates of these different scenarios are directly linked.}. 

\paragraph{SN ejecta–companion precursor.}
An additional type of precursor is possible in the natal-kick channel. If the compact object is born in a core-collapse SN, expanding ejecta can impact the companion before the first peri-center. The companion intercepts a fraction $\sim (R_\ast/2a)^2$ of the ejecta kinetic energy,
\begin{equation}
E_{\rm coll}\sim \frac{1}{8} M_{\rm ej}v_{\rm ej}^2 \left(\frac{R_\ast}{a}\right)^2.
\end{equation}
This energy is released over a time ranging from $t_{\rm sh}\sim R_\ast/v_{\rm ej}$ (if the velocity distribution around $v_{\rm ej}$ is narrow) to $t_{\rm sh}\sim a/v_{\rm ej}$ if one assumes a reasonable ejecta velocity spread, $\Delta v_{\rm ej}\sim v_{\rm ej}$ (here the spread in travel times to the star dominates over the travel time of a particular $v_{\rm ej}$ across the star). It will be channeled into a soft, quasi-thermal flash (UV/soft X-ray) with $kT\sim$ few$\times 0.1$\,keV and a duration of minutes–hours \citep{Kasen2010,KutsunaShigeyama2015,HiraiPodsiadlowskiYamada2018,HiraiSawaiYamada2014}. For $a\!\sim\!10^{12}$\,cm, $R_\ast\!\sim\!R_\odot$, $M_{\rm ej}\!\sim\!1$–$5\,M_\odot$, $v_{\rm ej}\!\sim\!10^4$\,km\,s$^{-1}$, the radiated energy can reach $\sim 10^{46}$–$10^{48}$\,erg. The emission may be somewhat anisotropic (a low-density cone around the line of centers), reducing its detectability for random sightlines. A soft, short precursor would therefore favor the natal-kick channel; a faint, non-thermal one favors a grazing first passage.

{\it Jet breakout:} In the natal-kick channel, the relativistic jet is launched soon after the SN. As such it is necessary to verify that the jet can break through the SN ejecta. This can be estimated using the jet's isotropic equivalent luminosity as compared to the mass outflow rate of the SN ejecta, as encapsulated by \citep{Matzner2003,Bromberg2011,BDPG2020}
\begin{equation}
    \tilde{L}=\frac{L_{\rm j,iso}\beta_{\rm ej}}{\dot{M}_{\rm ej}c^2}\approx 0.15\eta_{\gamma,-1}^{-1}v_{\rm ej,8.5}L_{\rm \gamma,50}M_{\rm ej,0.5}^{-1}t_{\rm w,day}
\end{equation}
where $v_{\rm ej},M_{\rm ej}$ are the velocity and mass of the SN ejecta, $L_{\gamma}$ is the isotropic-equivalent $\gamma$-ray luminosity. The jet head propagates at $\beta_h=(\beta_{\rm j}+\beta_{\rm ej}\tilde{L}^{-1/2})/(1+\tilde{L}^{-1/2})$, and for the parameters above we get $\beta_h\approx 0.28 [\beta_{\rm ej,8.5}L_{\rm \gamma,50}t_{\rm w,day}/(\eta_{\gamma,-1} M_{\rm ej,0.5})]^{1/2}$. Using the head velocity we can estimate the breakout time $t_{\rm b}\approx t_{\rm w}\beta_{\rm ej}/(\beta_{\rm h}-\beta_{\rm ej})\approx 3\times 10^3 (\beta_{\rm ej,8.5} t_{\rm w,day}M_{\rm ej,0.5}\eta_{\gamma,-1}/L_{\rm \gamma,50})^{1/2} \mbox{ s}$. Successful breakout requires an engine duration that is longer than $t_{\rm b}$. Indeed, the very long duration of GRB 250702B, satisfies this criterion for the canonical set of parameters considered here. At the same time, we see that the breakout is not trivial. This might, in practice, pose an important constraint for other $\mu$TDE systems and essentially acts as a selection effect guaranteeing that only particularly long-duration events of this class can result in a successful jet breakout.

\subsection{Hybrid Natal-kick + Partial Channel}
\label{subsec:hybrid}
A natural intermediate scenario is that the compact object is newly born and receives a natal kick that seeds a close, eccentric post-SN orbit, while the \emph{first} peri-center passage is only partial ($\beta\lesssim 1$). The stripped debris from this first, grazing encounter can power a faint pre-peak signal, while the bound stellar core returns after an orbital period of $\sim$day for a deeper encounter that powers the ultra-long main flare. Compared with the pure dynamical channel, the hybrid case inherits (i) the ballistic delay and possible soft, quasi-thermal SN–companion shock signature (Sec.~\ref{subsec:Kicks}) \emph{and} (ii) the multi-episode behavior and steeper late-time fallback slopes characteristic of partial disruptions (Sec.~\ref{subsec:partial}). 

\paragraph{Rates.} In both dynamical and natal-kick settings, the differential probability in the gravitational-focusing regime scales as $\mathrm{d}P/\mathrm{d}\ln r_p \propto r_p$. Since the ranges of $r_p$ corresponding to full and partial outcomes differ only by factors of order unity around $r_{\rm t}$, the partial and full-disruption \emph{cross sections are comparable} to within factors of a few.  Consequently, hybrid natal-kick + partial events should occur at a rate similar (within $\sim$factor-few) to the pure natal-kick full-disruption channel.

\subsection{Afterglow expectations and tests}
\label{subsec:afterglow}
Relativistic ejecta from a $\mu$TDE should produce a synchrotron afterglow that peaks once decelerated by the ambient medium. For `thin shells' the jet starts decelerating significantly once the the reverse shock crosses the jet material, which occurs at a source frame time
\begin{equation}
 t_{\rm dec,tn}\;\approx\;\left[\frac{(3-k)E_{\rm k,iso}}{2^{5-k}\pi n(R_0) R_0^k m_p c^{5-k} \Gamma_0^{8-2k}}\right]^{\frac{1}{3-k}}, 
\end{equation}
where $E_{\rm k,iso}$ is the isotropic-equivalent kinetic energy, $\Gamma_0$ the initial bulk Lorentz factor, $n(R_0)$ is the ambient density at some radius $R_0$ and the density is assumed to have a radial profile $n=n(R_0)(R/R_0)^{-k}$ with $0\leq k<3$. Following the deceleration time, the forward external shock produced by the interaction of the jet with the ambient medium follows a self-similar evolution which depends only on the local dynamics along the direction of motion (Blandford-Mckee phase, see \citealt{BM1976}), and leading to $\Gamma\propto t^{\frac{k-3}{8-2k}}$.
This self-similar phase lasts until $\Gamma\approx \theta_{\rm j}^{-1}$ or the jet break time $t_{\rm j}=t_{\rm dec,tn} (\Gamma_{0} \theta_{\rm j})^{\frac{8-2k}{3-k}}$, at which point the jetted nature of the outflow can no longer be ignored and the jet can start spreading laterally; $t_{\rm j}$ 
is characterized by a largely achromatic steepening of the afterglow lightcurve (known as a jet break).

From the $\gamma$-ray emission we have inferred $E_{\rm k,iso}\!\sim\!1.4\times10^{55}\eta_{\gamma,-1}^{-1}$\,erg. In addition, taking $n(10^{18}\mbox{ cm})\sim 1\mbox{ cm}^{-3}$ and a `wind' medium with $k=2$ (or more accurately, $k=1.60\pm0.17$) as inferred from afterglow modeling \citep{O'Connor2025,Granot2025}, we see that for $\Gamma_0\gtrsim 30$ (as supported independently by compactness arguments based on the prompt emission observations), $t_{\rm dec,tn}<T_{90}$, leading to `thick-shell' deceleration in which the ejecta decelerates before shock crossing and the energy extraction / emission from the external shock only peaks at the later shock crossing time, $\sim T_{90}$.
In a wind medium, the jet-break is very smooth and prolonged \citep{KP2000} and can be missed without a sufficient temporal baseline. The relatively steep decline of the X-rays, $F_X\propto t^{-1.8}$ from the first days and onward might indicate that the observations on timescales of days to tens of days are taking place in the midst of a jet break transition or even post-jet break, if $p\approx 2$ and the jet is non-spreading.
That being said, the decline is still potentially consistent with a pre-jet break temporal decline as long as the X-rays reside below the cooling break, $k\approx 2$ and the power-law index of the energy distribution of the accelerated electrons is $p\approx 2.6$. 
More generally, clearly identifying the jet-break from the multi-wavelength lightcurves will enable to measure $\theta_{\rm j}$, and in turn will provide us with improved estimates of the true collimated-corrected energy associated with the event. At later times (of order months to years), once the blast-wave has transitioned to non-relativistic velocities, the radio lightcurve can provide a more direct estimate of the outflow energy.

In the context of $\mu$TDEs, (as in other relativistic TDEs) there can be another source of X-rays beyond an external shock afterglow, which is due to direct emission from the accretion disk or internal activity of the jet. Known relativistic TDEs exhibit declining X-ray lightcurves which are initially highly variable and then shut-off abruptly on a timescale of a hundred to few hundred days \citep{Eftekhari2024}. Both the rapid variability and the very abrupt drop of the X-ray emission, imply that the observed X-rays cannot be produced in an external shock and indicate emission from smaller radii. The situation regarding GRB 250702B remains ambiguous. It is extremely variable in the first few days (indicating non-afterglow origin), but the X-ray decline on slightly longer timescales is consistent with the emission seen at longer wavelengths and is potentially consistent with an afterglow origin. No steep drop has been seen in the X-rays so far. With more data, multi-wavelength modeling can be used to critically test the afterglow scenario (e.g. by using closure relations between spectral and temporal slopes). At a minimum, afterglow emission should not over-produce the emission compared to what is detected. 

The interpretation of the X-ray afterglow must be considered in the context of the available multi-wavelength data. GRB 250702B was detected in X-rays across both the soft (Swift/XRT) and hard (NuSTAR) bands \citep{O'Connor2025}, while IR-optical and radio observations provide additional, though limited, constraints. The X-ray spectra are well described by non-thermal power-law,  or broken power-law spectral models. Such spectral properties are broadly consistent with synchrotron emission from a relativistic outflow. However, as discussed in \cite{O'Connor2025}, significant dust extinction along the line of sight—both in the Milky Way and in the host galaxy—introduces substantial uncertainty in the intrinsic optical-to-X-ray spectral slope, limiting our ability to tightly constrain the broadband spectral energy distribution at late times.
Moreover, the X-ray spectral index alone does not uniquely identify the emission mechanism. Relativistic tidal disruption events with steep X-ray light-curve declines, \citep{Seifina2017,Eftekhari2024}, exhibit non-thermal X-ray spectra comparable to those expected from standard afterglow synchrotron emission. As a result, while the broadband and X-ray spectral properties favor a non-thermal origin for the late-time emission, they do not by themselves distinguish between an external-shock afterglow and a long-lived relativistic internally dissipating jet powered by accretion. We therefore base our interpretation primarily on the combined temporal and energetic properties of the emission, while noting that deeper multi-band observations with improved extinction constraints would be required to definitively discriminate between these possibilities.

The dynamical and natal-kick $\mu$TDE channels lead to different afterglow signatures. The first is likely to correspond to either a uniform ISM ($k=0$) or a wind like medium due to the companion star. In the second case, surrounding material has been largely swept up by the SN ejecta. The jet must thus breakout of the SN ejecta before it starts decelerating significantly. As shown in \S \ref{subsec:Kicks}, this will only happen if the engine duration is extremely long - likely leading to thick shell deceleration in this case. After breaking out of the SN ejecta, the jet will likely encounter wind material that has been ejected from the star before its collapse, typically corresponding to $k\approx 2$. If the initial orbital period is much longer than a day, then the SN ejecta is optically thin to $\gamma$-rays, and the main phase of the GRB can be produced before the jet breaks out of the SN ejecta and does not require the engine to be particularly long lived. In this situation, the afterglow onset will be delayed and occur only once the jet has surpassed the SN ejecta material \citep{BP2024}.

\begingroup 
    \setlength{\tabcolsep}{6pt} 
    \renewcommand{\arraystretch}{1} 
    \setlength\extrarowheight{2pt}
    \begin{table*}
        \centering
        \caption{Channel diagnostics for a day-earlier pre-peak signal}
        \label{tab:diagnostics}
        \begin{tabular}{c c c c }  
            & \textbf{Dynamical (partial/repeating)} & \textbf{Natal kick (full)} & \textbf{Hybrid natal-kick + partial} \\
            \hline \hline     
            \textbf{pre-peak signal spectrum} & Non-thermal, faint/soft & Soft quasi-thermal & Both possible: soft quasi-thermal \\
             & (grazing strip) & (SN or SN ejecta-companion collision) &  and faint non-thermal \\
             \hline
            \textbf{Delay scaling} & $\sim P_{\rm orb}$& $\sim a/v' \sim P_{\rm orb}/2$& Initial delay $\sim P_{\rm orb}/2$ and \\  
             & (return time) & (ballistic / half final orbital period) & subsequent delays of $\sim P_{\rm orb}$ \\ \hline
            \textbf{Main-phase slope} & $t^{-5/3}$; can steepen to & $t^{-5/3}$ & As partial: $t^{-5/3}$, \\
             & if a core remains & canonical & possibly steepening toward $t^{-9/4}$ \\ \hline
            \textbf{Repetition} & Possible weaker repeats & No repeats expected & Possible weak repeats\\
             &  &  & if the core survives \\
            \hline
            \textbf{Host preference} & Viable even in & Typically requires recent & Typically requires recent  \\
             & early-type (clusters) & star formation & star formation (kick) \\
            \hline \hline  
        \end{tabular}  
    \end{table*}
\endgroup

\section{Alternative Explanations}

Several other progenitor channels have been proposed to account for ultra-long GRBs. We briefly assess whether they can explain the unusual phenomenology of GRB\,250702B. Table \ref{tab:modelcompare} provides a summary of the comparison between the models and the main observables of GRB 250702B.

\emph{Blue-supergiant collapsars.}  
The collapse of extended hydrogen-rich stars can, in principle, sustain accretion for $\gtrsim 10^4$ s, naturally producing ultra-long bursts \citep{Woosley2012,Levan2014,Perna2018}. Indeed, the ULGRB GRB\,111209A was associated with the luminous SN\,2011kl \citep{Greiner2015}. However, collapsar models do not naturally predict pre-peak signal emission starting $\sim$1 day before the main emission. Moreover, such collapsars should almost invariably produce bright Type II supernovae, potentially in tension with the non-detection here. 

\emph{Magnetar engines.}  
Spin-down of a newly born, highly magnetized neutron star can power
long-lived relativistic outflows \citep{Metzger2015,Gompertz2017}.
Magnetars engines can inject (collimated-corrected) energies of up to $\sim few \times 10^{52}\mbox{ erg}$ towards powering a relativistic jet (the upper limit obtained by magnetars undergoing fallback on short, $\sim 10$\,s timescales, see \citealt{Metzger2018}) and have been invoked to explain plateau-like X-ray light curves and extended emission in some GRBs (although see \citealt{BL2021}). A serious challenge of magnetar central engines is that even if energy can potentially be injected over a sufficiently long timescale to explain ULGRBs, the energy per baryon needs to remain high enough to allow the flow to reach ultra-relativistic velocities. This consideration prevents magnetar central engines from accounting for ULGRBs, unless they are undergoing fallback accretion \citep{BGM2017,Metzger2018}. In the latter scenario, proto-magnetar engines may even result in moderate precursor gaps (where the gap in emission corresponds to a phase with low magnetization due to the interplay between the proto-magnetar becoming optically thin to neutrinos and eventually getting baryon loaded by the accretion; \citealt{Metzger2018}). That being said, the timescale for this precursor and the gap are $\lesssim 10-100$\,s, orders of magnitude shorter than observed for GRB 250702B.
Moreover, as in the blue-supergiant case, magnetar models with initial spin period and magnetic field that are conducive to powering ultra-long GRBs, are also accompanied by very bright supernovae, powered by material launched with lower energy per baryon \citep{Margalit2018,Metzger2018}. 

\emph{White-dwarf--intermediate-mass black hole disruptions.}  
Disruption of a white dwarf by an IMBH has also been proposed as a channel for ultra-long, relatively soft bursts \citep{Krolik2011,MacLeod2016,Ioka2016} as well as specifically for GRB 250702B \citep{Levan2025,Eyles-Ferris2025}. 
In principle, different WD compositions (He, hybrid, CO, ONe) have different radii, but for all of them $R_*\ll R_\odot$, so for $M_\bullet\sim10^4$-$5\times10^4\,M_\odot$, as required by $\gamma$-ray variability constraints, the characteristic WD fallback and viscous timescales remain $t_{\min},t_{\rm acc}\sim 10$-$10^2$~s \citep{Granot2025}, far shorter than the $T_{90}\gtrsim 12$\,ks prompt emission of GRB\,250702B. Moreover, the deep encounters needed to strip enough mass to power the beaming-corrected $\gamma$-ray energy tend to enter the tidal detonation regime rather than producing a long-lived fallback stream \citep[e.g.][]{Rosswog2009,MacLeod2016,Kawana2018}. Shallower encounters that avoid detonation supply only modest bound mass and produce flares on WD dynamical timescales ($\sim10$-$10^2$\,s). In addition, such events are expected predominantly in galactic nuclei or dense clusters hosting massive black holes, and might be less likely in off-nuclear regions as observed here.
WD-IMBH models proposed specifically for GRB\,250702B \citep[e.g.][]{Eyles-Ferris2025} reproduce aspects of the X-ray behavior but do not simultaneously satisfy these combined timescale, detonation, and energetic constraints. Taken together, these arguments make a WD donor in an IMBH TDE scenario for GRB\,250702B disfavored in our view.

\emph{Main-sequence--intermediate-mass black hole disruptions.}
A variant on IMBH TDEs, involves a disruption of a main sequence star \citep{Granot2025}. In order for the timescales to match with those observed in GRB 250702B, one requires that $\beta\gg 1$ (and that dissipation is dominated at $r_{\rm p}$) such that $t_{\rm min}$ may be sufficiently reduced, despite $M_{\bullet}$ being much larger than in the $\mu$TDE case, $M_{\bullet}\sim 10^{3.5}-10^4M_{\odot}$. The early X-rays before the $\gamma$-ray peak may correspond to initial accretion, forming a jet with an initially large baryon loading. An attractive feature of this scenario, is that the density profile inferred from afterglow modeling matches the conditions expected from Bondi accretion onto a BH of this mass scale. This, however, hinges on the systemic velocity between the IMBH and the gas to be sufficiently low, and, considering the large observed offset of GRB 250702B, would likely require the event to be associated with a stellar cluster. The $\gamma$-ray variability and hard $\gamma$-ray photons (implying a large Lorentz factor) inferred for GRB 250702B, are marginally consistent with this picture, but comparison with jets from `typical' GRBs and relativistic TDEs, lead us to favor a stellar mass black hole. In particular, due to the small amount of accreted mass relative to that of the BH, the spin of an IMBH is practically unchanged by the accretion, and so a large jet efficiency hinges critically on the initial IMBH spin.

\emph{Helium star mergers.} \cite{Neights2025} discuss a scenario in which a BH - star binary evolves into a common envelope phase which eventually results in the tightening of the orbit and the black hole merging with the helium core \citep{2001ApJ...550..357Z}. The large angular momentum involved in the merger can potentially launch a strong jet, and the event duration will be determined by the accretion timescale which can naturally be $>10^4$\,s, depending on the mass and composition of the star. The association with a stellar BH, makes the observed variability, hard $\gamma$-rays and $E_{\gamma}/L_{\rm X,11 hr}$ more natural, as per the $\mu$TDE scenario. The relatively slow increase of the jet power, as the BH spins up has been suggested to explain the delay from engine onset (i.e. the soft X-ray pre-peak signal) to the main $\gamma$-ray emission.

In summary, while each of these models can explain some aspects of ultra-long GRBs, it is highly challenging for them to explain the combination of properties seen in GRB\,250702B. In particular, the day-earlier pre-peak signal, the off-nuclear location, the rapid variability and hard $\gamma$-ray emission and the absence of an obvious luminous supernova can be readily explained in the framework of a $\mu$TDE, while other suggested channels may face significant difficulties. 

\begingroup 
    \setlength{\tabcolsep}{2pt} 
    \renewcommand{\arraystretch}{1} 
    \setlength\extrarowheight{2pt}
    \begin{table*}
        \centering
        \caption{Comparison between different models suggested for GRB 250702B}
        \label{tab:modelcompare}
        \begin{tabular}{c c c c c c c}  
            & \textbf{MicroTDE (this work)} & \textbf{BSG collapsar} & \textbf{Magnetar engine} & \textbf{WD IMBH TDE} & \textbf{MS IMBH TDE} & \textbf{He star mergers}\\
            \hline \hline   
            \textbf{$\gtrsim \! 12$\,ks prompt $\gamma$-rays} & $\checkmark$ & marginal &$\checkmark$ & unexpected & $\checkmark$  & $\checkmark$ \\
              &  & (long free-fall time) &  (with fallback) & & ($t_{\rm min}$ with $\beta\sim3$)  
            &  \\
            \textbf{day early pre-peak signal} & $\checkmark$ & early cocoon? & unexpected & unexpected & $\checkmark$ & $\checkmark$  \\
             \textbf{$\sim \!0.5$\,s $\gamma$-ray variability} & $\checkmark$ & $\checkmark$ & $\checkmark$ & marginal ($\lesssim\!\frac{10r_g}{c}$) & marginal ($\lesssim\!\frac{10r_g}{c}$) & $\checkmark$ \\
             \textbf{$\gtrsim \! 10$\,MeV photons} & $\checkmark$ & $\checkmark$ & $\checkmark$ & unknown & marginal & $\checkmark$  \\
            \textbf{Off nuclear position} & $\checkmark$ & $\checkmark$ & $\checkmark$ & Low probability & $\checkmark$ & $\checkmark$  \\
            \textbf{$\gamma$-ray/X-ray ratio} & $\checkmark$ & $\checkmark$ & $\checkmark$ & unknown & unknown  & $\checkmark$  \\
            \textbf{$\sim \!0.5$\,ks X-ray variability} & $\checkmark$ & unknown & unexpected & $\checkmark$ & $\checkmark$ & unexpected \\
            \textbf{No detected SN} & $\checkmark$ & bright SN expected & bright SN expected & SN expected & $\checkmark$ & $\checkmark$  \\
            & (SN in kick channel) &  & & (WD detonation) & & (weak SN expected)\\
             \textbf{Comment} & & & & & $n(r)$ matches afterglow &   \\
            \hline \hline  
        \end{tabular}  
    \end{table*}
\endgroup

\section{Discussion}
Both the dynamical interaction and natal-kick $\mu$TDE channels provide natural explanations for the timescales and energetics in GRB 250702B, whereas canonical ULGRB models struggle to reproduce the same phenomenology. A summary of some diagnostics of $\mu$TDE path-ways is given in table \ref{tab:diagnostics}. 

\emph{Event rates.}  
The rarity of ULGRBs is broadly consistent with theoretical expectations for $\mu$TDEs. Population synthesis and dynamical studies suggest rates of $\sim 1$--$100~{\rm Gpc^{-3}\,yr^{-1}}$ for stellar-mass BH--star disruptions in clusters or binaries \citep{Perets2016,Kremer2019,Kremer2021,Kremer2023}, comparable to or somewhat above inferred ULGRB rates \citep{Levan2014,Levan2015}. As detailed in \S \ref{subsec:hybrid}, both the dynamical and natal kick channels result in comparable rates of partial and full TDEs. This leads to a third, `hybrid' channel, which involves natal-kicks leading to initial partial disruption and repeating deeper encounters. Overall, the three $\mu$TDE channels are expected to contribute at comparable levels within order-unity factors, consistent with the small observed sample size.

{\it Variability.} The association to a stellar black-hole can easily admit short variability times as observed in the $\gamma$-rays of GRB 250702B and other ULGRBs (note that 0.5 s is still 5000 times greater than the Schwarzschild crossing time of a $10M_{\odot}$ black-hole) and likely corresponds to emission from a relativistic jet. At the same time, $\mu$TDEs predict also longer timescale variability ($\sim 10^2$–$10^3$ s) during the early $\gamma$-rays and X-rays (as seen in the early X-rays of GRB 250702B), arising from stream–stream shocks and intermittent circularization near $R_{\rm circ}\!\simeq\!2r_p$, consistent with variability seen in GRB 250702B.

\emph{X-ray light-curve diagnostics.}  
Partial/repeating disruptions produce an early steep decline if time is measured from the first passage, followed by a $\sim t^{-5/3}$ tail anchored at the main encounter, and potentially a later steepening toward $t^{-9/4}$ if a bound core persists \citep{Coughlin2019,Wang2021}. The observed light curve of GRB\,250702B-steep when referenced to the pre-peak signal, but closer to $t^{-5/3}$ when measured from the main event-is consistent with this picture.

\emph{Predictions and near-term tests.}
(i) \emph{Jet break and shutoff:} A break in the multi-wavelength lightcurves at $t_j$ with post-break steepening tests the beaming angle, and constrains the energetics. If an abrupt shutoff of the X-ray emission is seen it would indicate that the preceding observed X-rays were internally produced and not due to the external shock (although the timescale associated with a transition to sub-Eddington accretion is orders of magnitude longer for stellar black holes than SMBHs).
(ii) \emph{Radio calorimetry:} A late-time ($\sim$year) flattening in radio decay in the trans-relativistic phase yields a beaming-independent energy, directly testing the disrupted-mass budget.
(iii) \emph{Spectral closure:} External-shock closure relations should hold only during the external shock afterglow phase.
(iv) \emph{Multi-episode prompt:} Recurrence (a third, weaker episode at $\sim$another orbital period) would strongly support partial/repeating; its absence is natural in the case of full disruption by natal-kicks.
(v) \emph{Polarization:} Moderate prompt linear polarization with angle swings would be consistent with a jet whose launch axis can precess under Lense–Thirring torques from a tilted debris disk.

\emph{Rates.}
Estimating $\mu$TDE rates is necessarily uncertain--the dominant formation channels are not well constrained, and the fraction of events that successfully launch relativistic jets (and their opening angles) is poorly known. Nevertheless, existing order-of-magnitude estimates suggest that $\mu$TDEs can plausibly occur at rates comparable to (or exceeding) the intrinsic ULGRB rate. Analytic estimates by \citet{Perets2016} give BH-star $\mu$TDE rates per Milky-Way-like galaxy, $\Gamma_{\mu{\rm TDE}}$, of
$\sim2.8\times10^{-6}\,\mathrm{yr^{-1}}$ from encounters in dense stellar systems and $\sim1.4\times10^{-6}\,\mathrm{yr^{-1}}$ from natal kicks in binaries, while perturbed wide binaries contribute a lower rate $\sim10^{-7}\,\mathrm{yr^{-1}}$ (their Table~2). Converting these to volumetric rates using a local galaxy number density
$n_{\rm gal}\sim(3$--$10)\times10^{-3}\,\mathrm{Mpc^{-3}}$ yields
\begin{equation}
\rho_{\mu{\rm TDE}}\sim \left( \frac{\Gamma_{\mu{\rm TDE}}}{\rm gal^{-1}\,yr^{-1}} \right)
\left( \frac{n_{\rm gal}}{\rm Mpc^{-3}} \right)\,10^{9}
\;\;{\rm Gpc^{-3}\,yr^{-1}},
\end{equation}
so that the \citet{Perets2016} dense--system and natal--kick channels correspond to $\rho_{\mu{\rm TDE}}\!\sim\! 8-28$ and $\sim4-14\,\mathrm{Gpc^{-3}\,yr^{-1}}$, respectively, whereas the wide--binary channel gives $\rho_{\mu{\rm TDE}}\sim0.3$--$1\,\mathrm{Gpc^{-3}\,yr^{-1}}$. More detailed dynamical modeling yields a local globular--cluster BH--MS TDE rate of order $\sim3\,\mathrm{Gpc^{-3}\,yr^{-1}}$ \citep{Kremer2019}, while simulations of young star clusters predict substantially higher values--up to $\lesssim200\,\mathrm{Gpc^{-3}\,yr^{-1}}$ \citep{Kremer2021}--and recent work suggests a total $\mu$TDE rate of $\sim350$--$450\,\mathrm{Gpc^{-3}\,yr^{-1}}$ at $z\simeq0$ (rising to $\sim2000$--$3000\,\mathrm{Gpc^{-3}\,yr^{-1}}$ at $z\simeq2$), dominated by repeated dynamical encounters, with SN--kick--induced $\mu$TDEs contributing only $\sim1$--$5\,\mathrm{Gpc^{-3}\,yr^{-1}}$ \citep{Rastello2025}.

For comparison, the inferred on-axis ULGRB rate is $\rho_{\rm obs}\approx0.1$--$0.6\,\mathrm{Gpc^{-3}\,yr^{-1}}$ \citep{Prajs2017}. The observable on--axis rate expected from $\mu$TDEs can be written as
\begin{equation}
\label{eq:rate}
\rho_{\rm obs}\sim \rho_{\mu{\rm TDE}}\, f_{\rm jet}\, f_b,
\end{equation}
where $f_{\rm jet}$ is the fraction of $\mu$TDEs that produce ULGRB--like relativistic jets and $f_b$ is the beaming correction. Using the small--angle approximation $f_b\simeq \theta_{\rm j}^2/2$. The inferred $\theta_{\rm j}$ for GRB 250702B is of order only a few degrees \citep{Granot2025}, while previous ULGRBs were inferred to have $\theta_{\rm j} \gtrsim 12^{\circ}$ due to the lack of observed jet breaks \citep{Prajs2017}.
Taking $\theta_{\rm j}\sim 3^\circ-12^\circ$ corresponds to $f_b\sim 1.4-22\times 10^{-3}$.
Thus, if $\rho_{\mu{\rm TDE}}\sim100\,\mathrm{Gpc^{-3}\,yr^{-1}}$ (within the range suggested for young--cluster dominated populations), then from Eq. \ref{eq:rate} we would infer $f_{\rm jet}\sim0.05-0.7$ which is physically plausible. The $\mu$TDE and ULGRB are therefore consistent within the level of uncertainty that currently exists in our determination of the different component in Eq. \ref{eq:rate}.

Finally, the hybrid natal-kick$+$partial disruption channel discussed in \S \ref{subsec:hybrid} is not expected to be strongly rate--suppressed relative to natal--kick full disruptions: in the gravitational--focusing regime the distribution of pericenter passages near $r_t$ implies that partial and full disruptions have comparable cross sections (within factors of a few), so the hybrid channel should occur at a similar order--of--magnitude rate.

\emph{Host environments.}  
The off-nuclear location disfavors models tied to galactic nuclei. $\mu$TDEs are expected both in stellar clusters (where BH--star encounters occur dynamically) and in massive binaries perturbed by natal kicks. Importantly, ULGRBs in \emph{early-type} (quiescent) galaxies would be \emph{unlikely} for collapsars (which require recent massive star formation) and are also disfavored for pure natal-kick events tied to core-collapse SNe. In contrast, the \emph{dynamical} channel remains viable in early-type hosts via interactions in dense, old stellar environments. Thus, a secure ULGRB in an early-type galaxy would strongly favor the dynamical (or hybrid with an old compact object) $\mu$TDE origin.

\emph{Broader implications.}  
If confirmed, GRB\,250702B would provide the first compelling evidence for $\mu$TDEs, extending the landscape of high-energy transients beyond the classical long/short GRB dichotomy. Such events probe stellar remnants interacting with ordinary stars on AU scales, opening a new window onto black hole demographics, natal kick physics, and stellar dynamics in dense environments. In addition, pre-peak signal flares may serve as powerful observational signposts of this channel, enabling early alerts for future ultra-long bursts with existing missions such as \textit{Einstein Probe} 
and future missions such as \textit{Theseus} \citep{Amati2021}.

Overall, while ULGRBs as a class likely comprise multiple progenitor channels, the unique features of GRB\,250702B are most consistently explained by a micro-TDE origin.

\section{Conclusions}

GRB\,250702B exhibited a unique combination of features:
an ultra-long duration of many hours, along with very rapid $\gamma$-ray variability, several MeV photons and a relation between isotropic $\gamma$-ray emission and X-rays detected $\sim 0.5$days later that is typical of different GRB classes, a pre-peak signal X-ray flare detected $\sim$1 day earlier, an off-nuclear position relative to its host galaxy, and no evidence for an accompanying supernova. These properties are difficult to explain within standard ULGRB models, such as blue-supergiant collapsars, magnetar engines, or white-dwarf disruptions by IMBHs.

We have shown that a micro-tidal disruption event ($\mu$TDE), the 
tidal stripping or disruption of a main-sequence star by a stellar-mass black hole or neutron star, provides a natural explanation. 
The $\mu$TDE may be a result of dynamical interactions or of natal kicks, and may involve partial/repeating disruption (in the former) or either partial/repeating or full disruption (in the latter). These channels can naturally reproduce a pre-peak signal–main flare separation, associated with the orbital period of the binary. The long emission timescale and energetics are consistent with fallback and viscous accretion in $\mu$TDEs, and the absence of a bright supernova is naturally accounted for.

If correct, this interpretation would constitute the first strong observational evidence for micro-TDEs, extending the diversity of high-energy transients beyond the standard GRB taxonomy. 
Future wide-field high-energy missions, particularly those with sensitive X-ray pre-peak signals will be critical for systematically identifying such events and constraining their rates, thereby probing compact-object dynamics and stellar interactions on AU scales.

\section*{Acknowledgments}
We thank Mike Moss, Brendan O'Connor, Eric Burns and Pawan Kumar for helpful discussions. PB's work was funded by grants (no. 2020747, 2024788) from the United States-Israel Binational Science Foundation (BSF), Jerusalem, Israel and by a grant (no. 1649/23) from the Israel Science Foundation.

\bibliographystyle{aasjournal}

\bibliography{MicroTDE}




\end{document}